\documentstyle[titlepage,12pt]{article}
\begin{document}
\title {Generation of the Scalar Field and Anisotropy \\ at Quantum Creation
of the Closed Universe }
\author { V.N.~Folomeev, 
V.Ts.~Gurovich \thanks{email: gurovich@grav.freenet.bishkek.su}
and I.V.~Tokareva\\
{\it Physics Institute, National Academy of Science, Kyrgyz Republic}}
\maketitle

\begin{abstract}
The behaviour of the wave function of the Universe under the barrier for anisotropic 
cosmological  Bianchi type IX model with account of influence of the scalar field 
is explored. In view of known difficulties with interpretation of multidimensional 
wave functions the method of reduction of such problems to one-dimensional is offered. 
For this purpose in frameworks of semiclassical approach the system of characteristics 
equations relative to one variable is written out. This system describe a bundle of 
the characteristics along which the multidimensional problem is reduced to one-dimensional 
one that allows to utillize the standard interpretation of the wave function as well as 
for usual Schr\"{o}dinger equation. The obtained  results for  Bianchi type IX model 
are reduced to the following statement: the Universe tunnels through the barrier from 
an isotropic state with zero initial value of the scalar field and appear in classically 
allowed region with small anisotropy that is necessary for providing of long-lived 
inflation for deriving the Universe such as ours.
\end{abstract}

\section{Introduction}

One of the basic problems of a classic cosmology is the presence of an
initial singularity. For its overcoming is considering the process of the
quantum creation of the Universe. Such approach was suggested by 
DeWitt~\cite{ref:DeWitt} and Misner~\cite{ref:Misner} and further was developed by
series of the authors. Ya. Zel'dovich~\cite{ref:Zeld} and A. Vilenkin~\cite{ref:Vil1}
have offered the mechanism of the spontaneous creation of the
Universe from ''nothing'' where ''nothing'' means an absence of classic
spacetime. The cosmological wave function can be used for calculation of
probability distribution of initial configurations of born Universes.

However, the quantum cosmology being based on quantum gravity maintains
all of its problems, in particular the inherent divergences. In addition
there are some other problems related to process of quantization of the
closed Universe. First of them follows from the fact that the wave function
of the Universe $\psi $\ does not depend on time. It should be understood
~\cite{ref:DeWitt, ref:Linde} in the sense that the wave function should describe
everything, including the clocks. In other words, time is a some interior
parameter intrinsical given configuration and expressed through the material
or geometrical variables. The important requirement imposed on this
parameter is the monotony of its change. The second problem is the difficulties
emergent at attempt of correct determination of a probability density
current. We could attempt using of the conservation law of this 
current~\cite{ref:DeWitt}
\begin{equation}
\label{eq1}
j=\frac{i}{2}\left( \psi ^{\ast }\nabla \psi -\psi \nabla \psi ^{\ast
}\right) ,\,\,\,\,\,\nabla j=0,
\end{equation}
but here we conflict with the same problem as in a case of the relativistic
Klein-Gordon equation: the probability density current obtained by this way is not
positive-definite. However, though we do not know how to solve these problems in
general it is possible to do at semiclassical statement of the problem~\cite{ref:Vil2}.

On the other hand, it is very difficult to find
physical interpretation of wave functions even in a semiclassical 
approximation in case of
minisuperspace with several degrees of freedom. Thereupon some
authors offers the methods of reduction of multidimensional problems to
one-dimensional~\cite{ref:Most, ref:Shau}.

The old problem that our Universe even in early youth ''did not want to be''
anisotropic and inhomogeneous is only partially resolved in models of the
early Universe. It is seems probable that the causality of the Universe at
the moment of creation allows to smooth out inhomogeneities of an energy
distribution. However, we have no such answer for anisotropy of the
Universe. The same problem is of interest for models of quantum
cosmology with creation of the Universe from ''nothing''. As is known, the
given scenario requires a finiteness of Euclidean action for the Universe
and consequently the Wheeler-DeWitt equation (WDW) for Bianchi
type IX models is explored. Closed Friedmann cosmological model is 
rather special case of this ensemble.

The relevant WDW equation for Bianchi type IX was considered in other statement of
problems earlier. In particular, Del Campo and Vilenkin calculated the wave
function of the Universe in some extreme cases when the anisotropy is either 
small or large~\cite{ref:Del}. Amsterdamski~\cite{ref:Amst}
calculated the wave function in semiclassical approximation at assumption
that anisotropy is small.

The peculiar behaviour of the wave function under the barrier clarified by the
authors in~\cite{ref:Gur} allows to consider the indicated problem in
another statement. In instantonic treatment the transition to imaginary time
is similar to considering of the WDW equation under the barrier. Thus, as is
noted in~\cite{ref:Gur} , equation for physical parameters, in
particular for the scalar field changes a sign of a kinetic energy under the
barrier. It gives in appearance of a principle opportunity of increase of the
field during tunneling. This feature was utilised for the scenario of
creation of the Universe from ''nothing'' with small initial energy density
of the scalar field. The quantity of last one starts to grow under the
barrier that gives on boundary of exterior classically allowed region
the value is necessary for sufficiently long period of inflation and
further transition to the standard hot Universe. The indicated
value of field is determined by the ''easiest'' way of tunneling.

The similar problem is considered for Bianchi type IX model in this paper.

\section{WDW equation for Bianchi type IX model}

The metric for Bianchi type IX has the following form (in this paper we use
Planck unities $c=G=\hbar =1$)~\cite{ref:Misner1}:
\begin{equation}
\label{eq2}
ds^{2}=dt^{2}-a^{2}e^{2\beta _{ij}}\sigma ^{i}\sigma ^{j},
\end{equation}
where $\sigma ^{i}$ represents the dual 1-forms satisfying to relations $%
d\sigma ^{i}=\eta _{jk}^{i}\sigma ^{j}\wedge \sigma ^{k}$ and
\begin{eqnarray}
\label{eq3}
\sigma ^{1} &=&\cos \psi d\theta +\sin \psi \sin \theta d\phi,  \nonumber \\
\sigma ^{2} &=&\sin \psi d\theta -\cos \psi \sin \theta d\phi, \\
\sigma ^{3} &=&d\psi +\cos \theta d\phi  \nonumber,
\end{eqnarray}
$0\leq \theta \leq \pi ,\,0\leq \phi \leq 2\pi ,\,0\leq \psi \leq 4\pi $,
and the diagonal tensor $\beta _{ij}$\ is parametrized as follows:
\begin{equation}
\label{eq4}
\beta _{ij}=diag\left( \beta _{+}+\sqrt{3}\beta _{-},\,\beta _{+}-\sqrt{3}%
\beta _{-},\,-2\beta _{+}\right) .
\end{equation}

As is known, the basic equation describing the quantum evolution of the
Universe is the WDW equation~\cite{ref:DeWitt}. For its construction we
shall consider the theory of the scalar field $\phi $\ with Lagrangian
\begin{equation}
\label{eq5}
L=-\frac{R}{2k}+\frac{1}{2}\left( \partial _{\mu }\varphi \right)
^{2}-V\left( \varphi \right) ,
\end{equation}
here $k=8\pi $\ is Einstein gravitational constant. From here, by taking
advantage (\ref{eq2}), (\ref{eq3}) and (\ref{eq4}) we obtain:
\begin{eqnarray}
\label{eq6}
L =6\pi ^{2}\left\{ \frac{a\dot {a}^{2}}{k}-\frac{a^{3}}{k}\left( 
\dot {\beta }_{+}^{2}+\dot {\beta }_{-}^{2}\right) +\frac{a}{k}%
\left( U\left[ \beta _{+},\beta _{-}\right] -1\right) -\frac{a^{3}}{3}\left[ 
\frac{1}{2}\dot {\varphi }^{2}-V\left( \varphi \right) \right]
\right\} , \\
U\left[ \beta _{+},\beta _{-}\right] =1+\frac{2}{3}e^{4\beta _{+}}\left[
\cosh \left( 4\sqrt{3}\beta _{-}\right) -1\right] +\frac{1}{3}e^{-8\beta
_{+}}-\frac{4}{3}e^{-2\beta _{+}}\cosh \left( 2\sqrt{3}\beta _{-}\right) , 
\nonumber
\end{eqnarray}
where through dot is designated derivative relative to time $t$. Further, we shall
find the momenta conjugate with $a,\varphi ,\beta _{+},\beta _{-}$:
\begin{equation}
\label{eq7}
p_{a}=\frac{12\pi ^{2}a\dot {a}}{k},\,\,p_{\varphi }=-2\pi ^{2}a^{3}%
\dot {\varphi} ,\,\,\,p_{\beta _{+}}=-\frac{12\pi ^{2}a^{3}\dot
{\beta} _{+}}{k},\,\,\,\,p_{\beta _{-}}=-\frac{12\pi ^{2}a^{3}\dot {%
\beta} _{-}}{k}.
\end{equation}
The Hamiltonian of the system is:
\begin{eqnarray}
\label{eq8}
H&=&-\frac{1}{4\pi ^{2}a^{3}}p_{\varphi }^{2}+\frac{k}{24\pi ^{2}a}p_{a}^{2}-%
\frac{k}{24\pi ^{2}a^{3}}\left( p_{\beta _{+}}^{2}+p_{\beta _{-}}^{2}\right)-\\
&-&6\pi ^{2}\left[ \frac{a}{k}\left( U\left[ \beta _{+},\beta _{-}\right]
-1\right) +\frac{a^{3}}{3}V\left( \varphi \right) \right]. 
\nonumber
\end{eqnarray}
Quantizing (\ref{eq7}) by replacement of momenta $p_{a},p_{\varphi },\,p_{\beta
_{+}},\,\,p_{\beta _{-}}$\ on the operators $-i\partial /\partial
a,$ $-i\partial /\partial \varphi ,\,-i\partial /\partial \beta
_{+},\,-i\partial /\partial \beta _{-}$ accordingly and using the rescaling $%
\varphi \longrightarrow \sqrt{3/\pi }/2\,\Phi $\ we write the WDW equation:
\begin{eqnarray}
\label{eq9}
-\frac{1}{a^{p}}\frac{\partial }{\partial a}\left(a^{p}\frac{\partial \Psi}{%
\partial a}\right)&+&\frac{1}{%
a^{2}}\left(\frac{\partial ^{2}\Psi}{\partial \Phi ^{2}}+
 \frac{\partial ^{2}\Psi}{\partial \beta _{+}^{2}}+\frac{\partial
^{2}\Psi}{\partial \beta _{-}^{2}}\right) +\\
&+&\frac{9}{4}\pi ^{2}a^{2}\left[
1-m^{2}a^{2}\Phi ^{2}-U\left[ \beta _{+},\beta _{-}\right] \right] 
\Psi=0.
\nonumber
\end{eqnarray}
In this equation potential of the scalar field is taken as $V\left( \varphi
\right) =m^{2}\varphi ^{2}/2$ where $m$\ is a mass of quantum of the scalar
field. So-called factor ordering $p$\ is also introduced. The usage of latter is
stipulated by ambiguity of commutation properties of $a$ and $p_{a}$\
~\cite{ref:Linde}. 

Transformation of the wave function
\[
\psi =a^{-p/2}\Psi 
\]
allows to eliminate first derivative in (\ref{eq9})
\begin{eqnarray}
\label{eq10}
-\frac{\partial ^{2}\psi}{\partial a^{2}}&+&\frac{1}{a^{2}}\frac{\partial
^{2}\psi}{\partial \Phi ^{2}}+\frac{1}{a^{2}}\left( \frac{\partial ^{2}\psi}{%
\partial \beta _{+}^{2}}+\frac{\partial ^{2}\psi}{\partial \beta _{-}^{2}}%
\right)-\\ 
&-&\frac{p}{2}\left( 1-\frac{p}{2}\right) \frac{1}{a^{2}}+\frac{9}{4}%
\pi ^{2}a^{2}\left[ 1-m^{2}a^{2}\Phi ^{2}-U\left[ \beta _{+},\beta
_{-}\right] \right]\psi =0,
\nonumber
\end{eqnarray}
or
\[
\left( \nabla ^{2}-W\right) \psi =0,
\]
where $\nabla ^{2}=\frac{1}{\sqrt{g}}\partial _{a}\left( \sqrt{g}g^{\alpha
\beta }\partial _{\beta }\right) ,\,\,\,g=\left| \det g_{\alpha \beta
}\right| $ and $g_{\alpha \beta }$\ is the ''metric'' in minisuperspace of
variables $a,\varphi ,\beta _{+},\beta _{-}$\ and the ''superpotential'' has
the form:
\begin{equation}
\label{eq11}
W\left[ a,\Phi ,\beta _{+},\beta _{-}\right] =-\frac{p}{2}\left( 1-\frac{p}{2%
}\right) \frac{1}{a^{2}}+\frac{9}{4}\pi ^{2}a^{2}\left[ 1-m^{2}a^{2}\Phi
^{2}-U\left[ \beta _{+},\beta _{-}\right] \right] .
\end{equation}

\section{Semiclassical approximation}

\subsection{Deriving of the system of characteristics equations}

Let's take advantage of the obtained WDW equation for examination of a
problem of quantum creation of the closed Universe from ''nothing''. For an
isotropic case such problem was considered in~\cite{ref:Gur}. As the
total energy of the closed Universe is zero it is clear from form of
superpotential (\ref{eq11}) that all space is parted into three regions: 

1)interior, representing classically unavailable area which we conventionally
call ''nothing''. In this region classical space and time does not exist that is similar 
to that there is no classical trajectory of a particle in forbidden area at $%
\alpha $- decay of atomic nucleus. The metric in this area experiences strong
quantum fluctuations. Its size is defined by the parameter of ordering $p$;

2) classically forbidden region under the barrier; 

3) classically allowed region. \\ 
The evolution of the wave function of the Universe represents
tunneling through the barrier from ''nothing'' and output in classically
allowed region. In this case the quantities of all physical parameters on which the
wave function depends are determined by process of tunneling instead of their
arbitrary choice on boundary of classically forbidden and allowed regions.

It is well known that the WDW equation (\ref{eq10}) has no the exact solution.
Therefore, we shall search for its approximate solution within the framework
of the semiclassical approach. Let's consider evolution of the wave function 
under the barrier. For this purpose we search for the solution of (\ref{eq10}) as $\psi
_{c}=e^{-S}$. The relevant equation for action $S\left( a,\varphi ,\beta
_{+},\beta _{-}\right) $\ shall be:
\begin{equation}
\label{eq12}
-\left( \frac{\partial S}{\partial a}\right) ^{2}+\frac{1}{a^{2}}\left[
\left( \frac{\partial S}{\partial \Phi }\right) ^{2}+\left( \frac{\partial S%
}{\partial \beta _{+}}\right) ^{2}+\left( \frac{\partial S}{\partial \beta
_{-}}\right) ^{2}\right] +W\left[ a,\varphi ,\beta _{+},\beta _{-}\right] =0.
\end{equation}
This equation represents an analog of the Hamilton-Jacobi equation from
classical mechanics. For finding of the solution of this nonlinear
differential equation it is possible to reduce it to system of the ordinary
differential equations called characteristic system of the given partial
equation. Utillizing this system it is possible to construct an integrated
surface of equation (\ref{eq12}) consisting from the characteristics. The required
system of the characteristics written relative to arbitrary parameter has a
form:
\begin{eqnarray}
\label{eq13}
\frac{d\Phi }{da} &=&-\frac{q}{a^{2}F},\,\frac{d\beta _{+}}{da}=-\frac{v}{%
a^{2}F},\,\frac{d\beta _{-}}{da}=-\frac{w}{a^{2}F},  \nonumber \\
\frac{dq}{da} &=&-\frac{9}{4}\frac{\pi ^{2}m^{2}a^{4}\Phi }{F},\,\frac{dv}{da%
}=-\frac{9}{4}\frac{\pi ^{2}a^{2}\partial U\left[ \beta _{+},\beta
_{-}\right] /\partial \beta _{+}}{2F},\\
\frac{dw}{da}&=&-\frac{9}{4}\frac{\pi
^{2}a^{2}\partial U\left[ \beta _{+},\beta _{-}\right] /\partial \beta _{-}}{%
2F}, \nonumber \\
\frac{dS}{da} &=&\frac{-\frac{1}{4a^{2}}+\frac{9}{4}\pi ^{2}a^{2}\left[
1-m^{2}a^{2}\Phi ^{2}-U\left[ \beta _{+},\beta _{-}\right] \right] }{F}. 
\nonumber
\end{eqnarray}
Here the designations are introduced:
\begin{eqnarray*}
q &=&\partial S/\partial \Phi ,\,\,v=\partial S/\partial \beta
_{+},\,w=\partial S/\partial \beta _{-}, \\
F &=&\sqrt{\frac{1}{a^{2}}\left( q^{2}+v^{2}+w^{2}\right) +\left[ -\frac{1}{%
4a^{2}}+\frac{9}{4}\pi ^{2}a^{2}\left[ 1-m^{2}a^{2}\Phi ^{2}-U\left[ \beta
_{+},\beta _{-}\right] \right] \right] ,}
\end{eqnarray*}
and the quantities $q,\,v,\,w$\ represent the ''momenta'' of the given
system and factor ordering $p=1$. In this case, the role of an arbitrary parameter 
is played by the scale factor $a$. The obtained system of equations describes
an one-dimensional motion of a ''particle'' along a characteristic. In
this case, monotonically varying parameter $a$\ can play a role of time
~\cite{ref:Vil3} to which there is an evolution of the Universe.
As the momentum of a ''particle'' is equal to a gradient of action, its
possible trajectories are orthogonal to surfaces $S=const$, i.e. surfaces
of a constant phase of the wave function. Thus, at transition to a classical
mechanics ''beams'' associated with function $\psi _{c}$\ (the trajectories
orthogonal to surfaces of a constant phase) represent possible trajectories
of a motion of ''particle''~\cite{ref:Vil2, ref:Schiff} .

\subsection{The analysis of the system of the characteristics}

Let's examinate directly the system of the characteristics
(\ref{eq13}). In view of equality to zero of a total energy of the Universe the
boundary between classically forbidden and allowed regions is defined from
requirement $W\left[ a,\Phi ,\beta _{+},\beta _{-}\right] =0$. At $\Phi
,\beta _{+},\beta _{-}=0$\ the Universe can exist infinitely long time in
the interior area posed in an interval
\begin{equation}
\label{eq14}
0\leq a\leq a_{0},
\end{equation}
where $a_{0}=\left( 1/9\pi ^{2}\right) ^{1/4}$\ .

For finding of the solutions of system of the characteristics (\ref{eq13}) it is
necessary to set boundary conditions which look like:
\begin{equation}
\label{eq15}
a=a_{0},\,\Phi =0,\,\beta _{+}=0,\,\beta
_{-}=0,\,q=q_{0},\,v=v_{0},\,w=w_{0},\,S=0.
\end{equation}
Let's specify that we shall start from a limiting isotropic state that
follows from equality to zero of parameters of an anisotropy $\beta _{+}$\ and $%
\beta _{-}$. Requirements on $q,v,w$\ means that  a
bundle of the characteristics with different initial values of ''momenta'' 
transits under the barrier.
The given bundle of the characteristics is restricted by the
upper limiting characteristic defined as $q_{0},v_{0},w_{0}\longrightarrow
-\infty $ and ground one which is coincident  with axis $a$\ 
at $q_{0},v_{0},w_{0}\longrightarrow 0$ (fig. 1). 
We are finding the values of quantities $a_{\ast },\Phi
_{\ast },\beta _{+\ast },\beta _{-\ast },\,S_{\ast }$ at an output of each 
characteristic on boundary
of the exterior classically allowed area $a=a_{\ast }$\ (the last is defined from 
requirement $W\left[ a_{\ast },\Phi _{\ast },\beta _{+\ast },\beta _{-\ast
}\right] =0$) .

It is clear from numerical analysis of system of the characteristics
 (\ref{eq13})  that the quantity of action $S$\ is great at small values of 
initial ''momenta'' $q,v,w$. The relevant penetrability of the barrier for such
characteristics will be exponentially small since last one will be defined
from
\begin{equation}
\label{eq16}
D=\exp (-2S).
\end{equation}
Therefore, the basic contribution to penetrability of the barrier will be
given by the characteristics for which $\left| q,v,w\right| >10$. On
the other hand, after escaping under the barrier a necessary stage
of the evolution of the Universe is
inflationary stage. For latter guarantee  two
important requirements are imposed on scalar field $\Phi $~\cite{ref:Linde}:
 1) value of the field
should be about Planck; 2) it should vary enough slowly with the purpose of
guarantee of a long-lived period of inflation that is necessary for an
stretching of the size of the Universe from Planck to macroscopic. For
sufficing these requirements it is necessary that the initial $q$\ were at list
 on the order more than $v$\ and $w$. Otherwise, anisotropy suppress the increase
of the scalar field and the requirements of guarantee of inflation are
disturbed.

It is possible significantly simplify  system of the characteristics (\ref{eq13})
based on condition of large value of parameters $q,v,w$\. We note that
in this case quantities $q,v,w$\ remains practically constant with
increasing of $a$. Then (\ref{eq13}) accept form:
\begin{eqnarray}
\label{eq17}
\frac{d\Phi }{da} &=&-\frac{q}{a^{2}F},\,\frac{d\beta _{+}}{da}=-\frac{v}{%
a^{2}F},\,\frac{d\beta _{-}}{da}=-\frac{w}{a^{2}F},  \nonumber \\
q &=&const,\,\,v=const,\,\,w=const, \\
\frac{dS}{da} &=&\frac{-\frac{1}{4a^{2}}+\frac{9}{4}\pi ^{2}a^{2}\left[
1-m^{2}a^{2}\Phi ^{2}-U\left[ \beta _{+},\beta _{-}\right] \right] }{F}, 
\nonumber
\end{eqnarray}
where $F\approx \sqrt{\left( q^{2}+v^{2}+w^{2}\right) /a^{2}}%
=A/a,\,\,\,A^{2}=q^{2}+v^{2}+w^{2}$. From (\ref{eq15}) we have:
\begin{equation}
\label{eq18}
\Phi =\frac{q}{A}\ln \left( \frac{a}{a_{0}}\right) ,\,\,\beta _{+}=\frac{v}{A%
}\ln \left( \frac{a}{a_{0}}\right) ,\,\,\beta _{-}=\frac{w}{A}\ln \left( 
\frac{a}{a_{0}}\right) ,
\end{equation}
\begin{equation}
\label{eq19}
\frac{dS}{da}=\frac{1}{A}\left[ -\frac{1}{4a}+\frac{9}{4}\pi ^{2}a^{3}\left[
1-m^{2}a^{2}\Phi ^{2}-U\left[ \beta _{+},\beta _{-}\right] \right] \right].
\end{equation}
For calculation of action S\ we shall note that the quantities of parameters
of an anisotropy remain always rather small (see fig. 1) that enables to
expand potential $U\left[ \beta _{+},\beta _{-}\right] $\ in series whence,
utillizing (\ref{eq18}) we have
\begin{equation}
\label{eq20}
U\left[ \beta _{+},\beta _{-}\right] =8\left( \beta _{+}^{2}+\beta
_{-}^{2}\right) =\frac{8}{A^{2}}\ln ^{2}\left( \frac{a}{a_{0}}\right) \left(
v^{2}+w^{2}\right) .
\end{equation}
Utillizing the last expression integrate (\ref{eq19}) that yields
\begin{eqnarray}
\label{eq21}
&& \qquad S=\int\limits_{a_{0}}^{a_{\ast }}\frac{1}{A}\left[ -\frac{1}{4a}+\frac{9}{%
4}\pi ^{2}a^{3}\left[ 1-m^{2}a^{2}\Phi ^{2}-U\left[ \beta _{+},\beta
_{-}\right] \right] \right] =  \\
&=&\frac{\pi ^{2}}{48A^{3}}-\frac{12A^{2}}{\pi ^{2}}\ln \left( \frac{a}{a_{0}}%
\right) +27q^{2}\left( a_{\ast }^{4}-a_{0}^{4}\right) -m^{2}q^{2}\left(
a_{\ast }^{6}-a_{0}^{6}\right) + \nonumber \\
&+&6a_{\ast }^{4}\left[ a_{\ast }^{2}m^{2}q^{2}+18\left( v^{2}+w^{2}\right)
\right] \ln \left( \frac{a}{a_{0}}\right) -18a_{\ast }^{4}\left[ a_{\ast
}^{2}m^{2}q^{2}+12\left( v^{2}+w^{2}\right) \right] \ln ^{2}\left( \frac{a}{%
a_{0}}\right).  \nonumber
\end{eqnarray}
As  it was already noted, the given formula usables only at large values $%
q,v,w $. But thus  it is visible from (\ref{eq21}) that the quantity of action is
small and accordingly evaluation of a penetrability of the barrier with help
of (\ref{eq16}) is impossible as the precision of WKB-approach is considerably
decreased. Therefore, in this case it is necessary to use the formula of
generalized WKB for evaluation of penetrability of the barrier~\cite{ref:Vil3} :
\begin{equation}
\label{eq22}
D=\frac{\exp \left( -2S\right) }{1+\exp \left( -2S\right) }.
\end{equation}
For calculation of average penetrability of the barrier it is necessary to
integrate (\ref{eq22}) on all $q,\,v$\ and $w$. Utillizing (\ref{eq21}) we are obtaining:
\begin{equation}
\label{eq23}
D=\lim_{\overline{q},\overline{v},\overline{w}\rightarrow \infty }
\frac{1}{\overline{q}\, \overline{v}\, \overline{w}}\int\limits_{0}^{%
\overline{q}\,}\int\limits_{0}^{\overline{v}\,}\int\limits_{0}^{\overline{w}%
\,}\frac{\exp \left( -2S\left( q,v,w\right) \right) }{1+\exp \left(
-2S\left( q,v,w\right) \right) }dqdvdw=\frac{1}{2}.
\end{equation}

\section{Conclusions}

The solution of a WDW equation is
inevitably conjugates to difficulties of interpretation of physical sense of
the wave function  even in a semiclassical approximation. 
Similarly to Klein-Gordon equation, even for two variables
such wave functions have no positive definiteness of the probability
distribution. If in the case of the Klein-Gordon equation
 this problem is solved because of
existence of antiparticles, so for the WDW equation this problem remains
unclosed. In this case,  the unique intelligent variant is the one-dimensional
treatment of the indicated equation with positive definiteness of the
probability distribution. This difficulty was mentioned by many authors
(see ~\cite{ref:Vil2} and references inside) at exceeding the limits of the
one-dimensional solutions. The consideration of the multidimensional WDW
solutions is inevitably connect with using of set of the one-dimensional
solutions. Problems numbered in the given paper is inevitably exceed the
limits of frameworks of the one-dimensional solutions. In this connection,
for tunneling of the wave function under the barrier in semiclassical
approximation we use the relevant characteristics equations permitting to
analyse the problem on a bundle of the characteristics with the subsequent
averaging of results on a selected bundle. Because of the real of wave
functions under the barrier such procedure doesn't conjugate to an interference
of wave functions. It does not give in interference problems during an
averaging on what the obtained above results are based.

Usage of the characteristics allows to avoid one more difficulty associated with
interpretation of multidimensional wave functions. It is often use an
analogy between the WDW equation (\ref{eq10}) and Schr\"{o}dinger equation.
 However,
there is an essential difference between them: in the Schr\"{o}dinger equation
the kinetic energy $-\nabla ^{2}$ is positive definite quantity and region
where $W>0$\ is really classically forbidden. In the WDW equation, generally
speaking, the operator $-\nabla ^{2}$\ is non-positive definite that is
associated with critical difference of signs of second derivatives. Therefore the
classic trajectories can penetrate ''classically forbidden'' region with
non-zero values of corresponding ''momenta''. In our formulation this
problem is solved by a natural way: at using of the characteristics the
problem becomes one-dimensional and any problems with signs of ''momenta''
do not arise as there is a unique momentum $dS/da$\ from (\ref{eq13}) along the
characteristic.

In the interior classically forbidden region (''nothing''), 
as well as in~\cite{ref:Gur}, 
the existence of the wave function of the Universe
relevant to a homogeneous isotropic Universe is supposed. Similar to~\cite{ref:Gur}
 during tunneling under the barrier the opportunity of increase of the
scalar field together with parameters of an anisotropy of model is supposed.
The relevant analysis of such scenario, shown above, displays that the
large anisotropy of the model collected under the barrier does not ensure
necessary increasing of an energy of the scalar field and, therefore, the
enough long-lived  inflation in the early Universe. 
The density of energy of the scalar field generated under the barrier 
which is sufficient for deriving of the Universe such as ours 
automatically ensures a small anisotropy at
transition of the wave function of the Universe in classically allowed
region. Thus, the result of the carried out examination is the following
statement: the cosmological model created from ''nothing'' from as much as
possible symmetric state (see, e.g.~\cite{ref:Vil4}) and ensuring the  
inflation which is requisite for deriving of the Universe such as ours
  has a small anisotropy with necessity.

This result, in principle, is in the consent with~\cite{ref:Yok} where
the problem on naturalness of inflation in Bianchi type IX model with
positive cosmological constant is explored. There it was shown  that in case
of small parameters of an anisotropy the inflation is inevitable otherwise
it is not obvious.

\par\bigskip
{\bf Acknowledgments}
\par\bigskip

\noindent 
We are grateful to R. Ruffini, V. Gurzadyan and A. Starobinsky for 
useful discussions of results.
This work was supported by the research grant~KR-154 of International 
Science and Technology Centre (ISTC).

\end{document}